\documentclass[%
 reprint,
 amsmath,amssymb,
 aps,
 superscriptaddress,
]{revtex4-2}

\usepackage{graphicx}% Include figure files
\usepackage{bm}% bold math
\usepackage{braket}
\usepackage{upgreek}
\usepackage{float}
\usepackage[dvipsnames,svgnames,x11names]{xcolor}
\usepackage{ulem}
\usepackage[colorinlistoftodos,prependcaption,textsize=tiny]{todonotes} % in-text comments

\usepackage{comment}

\usepackage[pdftex]{hyperref}
\hypersetup{
    breaklinks=true,
    colorlinks=true,
    linkcolor=MidnightBlue,
    urlcolor=NavyBlue,
    citecolor=Blue,
    pdftitle={Infrared spectroscopy of phase transitions in the lowest Landau levels of bilayer graphene}
}

\begin{document}

\newcommand{\eh}[1]{\todo[color=orange!60]{\footnotesize #1 }}
\newcommand{\ehi}[1]{\todo[inline,color=orange!60]{\footnotesize #1 }}

%\preprint{APS/123-QED}

\title{Infrared spectroscopy of phase transitions in the lowest Landau levels of bilayer graphene}

\author{B.\ Jordan Russell\textsuperscript{*}}
\affiliation{Department of Physics, Washington University in St. Louis, 1 Brookings Dr., St. Louis MO 63130, USA}
\affiliation{Center for Quantum Leaps, Washington University in St. Louis, 1 Brookings Dr., St. Louis MO 63130, USA}
\altaffiliation{These authors contributed equally to this work}
\author{Matheus Schossler\textsuperscript{*}}
\affiliation{Department of Physics, Washington University in St. Louis, 1 Brookings Dr., St. Louis MO 63130, USA}
\author{Jesse Balgley}
\affiliation{Department of Physics, Washington University in St. Louis, 1 Brookings Dr., St. Louis MO 63130, USA}
\author{Yashika Kapoor}
\affiliation{Department of Physics, Washington University in St. Louis, 1 Brookings Dr., St. Louis MO 63130, USA}
\author{T. Taniguchi}
\affiliation{National Institute for Materials Science, 1-2-1 Sengen, Tsukuba, Ibaraki 305-0044, Japan}
\author{K. Watanabe}
\affiliation{National Institute for Materials Science, 1-2-1 Sengen, Tsukuba, Ibaraki 305-0044, Japan}
\author{Alexander Seidel}
\affiliation{Department of Physics, Washington University in St. Louis, 1 Brookings Dr., St. Louis MO 63130, USA}
\author{Yafis Barlas}
\affiliation{Department of Physics, University of Nevada, Reno, 1664 N. Virginia Street, Reno, NV 89557}
\author{Erik A. Henriksen}
\affiliation{Department of Physics, Washington University in St. Louis, 1 Brookings Dr., St. Louis MO 63130, USA}
\affiliation{Institute of Materials Science \& Engineering, Washington University in St. Louis, 1 Brookings Dr., St. Louis MO 63130, USA}
\email{henriksen@wustl.edu}

\date{\today}

\begin{abstract}
We perform infrared magneto-spectroscopy of Landau level (LL) transitions in dual-gated bilayer graphene. At $\nu=4$ when the zeroth LL (octet) is filled, two resonances are observed indicating the opening of a gap. At $\nu=0$ when the octet is half-filled, multiple resonances disperse non-monotonically with increasing displacement field, $D$, perpendicular to the sheet, showing a phase transition at modest displacement fields from a canted anti-ferromagnet (CAFM) to the layer-polarized state, with a gap that opens linearly in $D$. When $D=0$ and $\nu$ is varied, resonances at $\pm\nu$ show an electron-hole asymmetry with multiple line splittings as the octet is progressively filled. The $\nu=4$ data show good agreement with predictions from a mean-field Hartree-Fock calculation when accounting for multiple tight-binding terms in a four-band model of bilayer graphene. However even by incorporating a valley interaction anisotropy tuned to the CAFM ground state, only partial agreement is found at $\nu=0$. Our results suggest additional physics is required to understand bilayer graphene at half-filling.
\end{abstract}

\maketitle

When neutral bilayer graphene is placed in a quantizing magnetic field, electron-electron interactions within the quartet of nearly-degenerate states of two Landau levels in combination with an electrostatically tunable energy difference between the layers lead to a rich phase diagram \cite{PhysRevLett.96.086805,McCann2013,Barlas2008,Kim2011a,Jung2011,Kharitonov2012,Lambert2013,Knothe2016,DeNova2017,Murthy2017,Green2020}. This has proven a fertile playground for the observation of competing, interaction-driven ground states and symmetry-breaking effects explored by a wide range of techniques including electronic transport \cite{Zhao2010,Weitz2010,Velasco2012,bao2012,Maher2013,Lee2014,Maher2014,Li2018,Li2019,Fu2021}, compressibility \cite{Henriksen2010a,Young2012,Kou2014,Hunt2017,Zibrov2017}, and scanning tunneling \cite{Rutter2011,Yin2020a} measurements. 

Spectroscopic measurements of excitations between Landau levels (LL) in the regime of these intriguing many-particle states are notably lacking among prior explorations, but can in principle yield insight into the nature of electronic interactions. In monolayer graphene these optical transitions are known to probe contributions from interactions beyond the single-particle inter-LL transitions \cite{iyengar_excitations_2007,jiang_infrared_2007,bychkov_magnetoplasmon_2008,Shizuya2010,Russell2018,Sokolik2019,Pack2020}. The same is widely anticipated for bilayer graphene \cite{Bisti2011,Shizuya2011,Toke2013a,Shizuya2020}, but the experimental literature is sparse \cite{Henriksen2008,Orlita2011,Moriya2021}. The phase diagram of bilayer graphene is best studied in dual-gated devices that allow independent control of the charge density and electric displacement field. Measurements based on scanned probes are ill-suited for this purpose, making optical approaches especially relevant as they can probe through infrared-transparent gates.

Recent scanning tunneling experiments in monolayer graphene sheet found a bond-ordered ground state at LL filling factor $\nu= h n / e B = 0$ ($n$ the zero-field carrier density, $-e$ the electron charge, and $B$ the magnetic field), at odds with the canted antiferromagnet (CAFM) state previously inferred from high field transport \cite{liu_visualizing_2022,Young2014}. Although the CAFM in \textit{bilayer} graphene is well-established, probing excitations in the bulk of the bilayer sheet directly may nonetheless be illuminating. We address this with a study of the far-infrared (ir) cyclotron resonance (CR) in a boron nitride-encapsulated, dual-gated graphene bilayer, focusing on excited state transitions to or from the quasi-zero energy Landau level octet. Our findings, consistent with the presence of a CAFM, reveal an intriguing complexity in the excited state transitions at $\nu=0$, and a surprisingly sharp phase transition from the CAFM to the fully layer (valley) polarized regime. 

\begin{figure*}[t!]
\includegraphics[width=\textwidth]{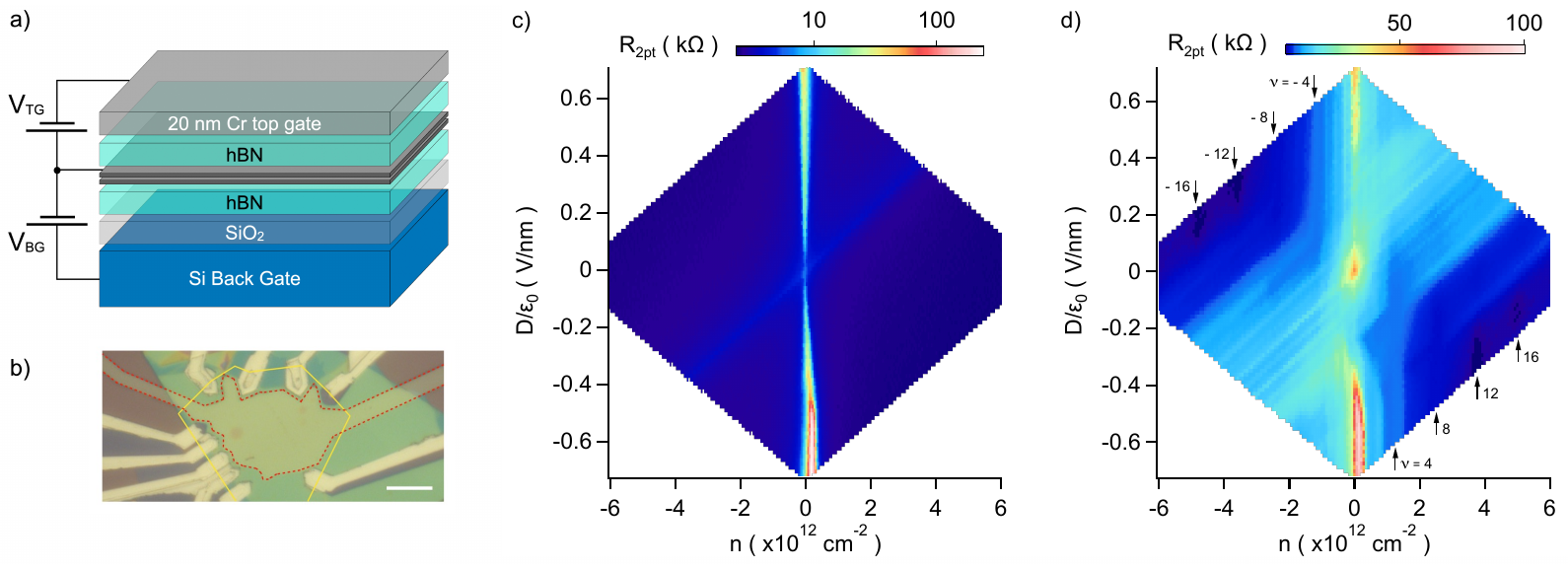}
\caption{(a) Schematic of dual-gated bilayer graphene device. (b) Optical micrograph taken at $50{\times}$ magnification, showing the $\sim$550 $\mu \text{m}^{2}$, 20-nm-thick ir-transparent Cr top gate (red outline) above the bilayer flake (yellow outline) encapsulated in the light green hexagonal boron nitride. Scale bar is 10 $\mu$m. (c) Two-terminal resistance at $B = 0$ T and $T=4$ K. (d) Two-terminal magnetoresistance at $B = 13$ T and $T\approx4$ K, with quantization features (vertical striping) at integer fillings between $\nu = \pm 4$. Diagonal features in (c) and (d) arise from back-gate bilayer graphene extending outside the top gate.  \label{transport}} 
\end{figure*}

Specifically, we measure CR transitions between the octet LLs ($N=0,1$) and $N = \pm 2$, for both $\nu = 0$ and $+4$, at a fixed magnetic field $B=13$ T (chosen to match a window of maximum far-ir transmission) while tuning the layer symmetry breaking displacement field, $D$. At $\nu = 4$ we observe the valley splitting within the $N = 1$ LL, which closely tracks the electrostatically tunable band gap at $B = 0$. At $\nu = 0$ we see an interaction-driven gapped state at $D = 0$, signatures of a spin-ordered ground state at low $D$, and a sudden sharp transition to the layer-polarized ground state at intermediate $D$. Finally, in holding $D=0$ fixed while tuning the filling factor over the range $\nu = \pm 6$, the CR line shapes show a weak electron-hole asymmetry with multiple line splittings and a pronounced increase in the transition energies right at $\nu = 0$.

To better understand these findings, we undertook an in-depth theoretical calculation of the inter-LL energies corresponding to these CR transitions \cite{SchosslerBLGfuture}. We progressively increased the order of approximation until the theory attained a very close match to the $\nu=4$ data and a reasonable if not perfect match to data at partial LL fillings. The single-particle physics is first calculated non-perturbatively for the four-band Slonczewski-Weiss-McClure (SWMc) tight-binding model in a $B$ field, accounting for both the energy difference $\Delta'$ between dimer and non-dimer sites, and the skew interlayer couplings $\gamma_{3}$ and $\gamma_{4}$ \cite{McCann2013,Zhang2011}. While the non-interacting spectrum qualitatively reproduced the phenomenology of the CR transitions at integer filling ($\nu = 4$), it fails to describe the results at partial filling of the octet.  This is expected as transition from the symmetry-broken states significantly influences the CR spectrum.

The effect of interactions was studied using a self-consistent Hartree-Fock mean field approximation including all four bands of the SWMc tight-binding model of bilayer graphene. Unlike previous studies \cite{Barlas2008,Toke2013a,Murthy2017,Shizuya2020}, we numerically project on to the $N=0,1$ octet and $\pm2$ levels, and then incorporate interactions comprising both the long-range layer (valley) independent Coulomb, and layer-dependent short-ranged interactions, which break the SU(4) symmetry of the octet. The inclusion of layer and valley anisotropy is essential to capture the broken symmetry phases at partial filling of the octet at $\nu=0$ \cite{Kharitonov2012}. In our calculations, we also include the influence of the infinitely deep sea of filled states on the zeroth LL \cite{Shizuya2011,Shizuya2020}, the Zeeman interaction, and the externally applied $D$ field. The energy bands are calculated self-consistently with a screening factor and, at $\nu=0$, the short-ranged interaction couplings are taken as adjustable parameters. As described below, a coherent picture consistent with the experimental data emerges when the full model described above is employed. Details of the theoretical calculations will be presented in a companion publication \cite{SchosslerBLGfuture}, while here we focus on the experimental phenomenology.

Both low-frequency electronic transport and far-ir transmission data were measured in a dual-gated, boron-nitride encapsulated bilayer graphene device with a global Si back gate and a $550~\mu$m$^2$, 20-nm-thick semi-transparent Cr top gate, all at a fixed magnetic field $B=13$ T and sample temperature $T\approx4$ K. Transmission spectra with a resolution of 1 meV were acquired at given $\nu$ and $D$ values, averaged for 4-8 hours, and normalized by spectra at different top gate voltages \cite{Pack2020}.

The electronic transport shown in Fig.\ \ref{transport} is consistent with prior observations, having a large resistance toward high $|D|$ values due to the tunable band gap, and another localized resistance peak near $\{n,D\}=0$ attributed to a CAFM quantum Hall insulator \cite{Maher2013}. Quantized transport is seen at all integer filling factors $|\nu| \leq 4$. As the top gate in the device shown in Fig.\ \ref{transport}(b) does not completely cover the bilayer sheet, the transport data in Fig.\ \ref{transport}(c,d) are contaminated by diagonal traces due to regions of the bilayer that only respond to the back gate. This issue does \textit{not} affect the infrared data: each spectra is normalized by other traces acquired at the same back gate but different top gate voltages, so that absorption features from the back-gated-only region cancel out.

\begin{figure*}[t]
\includegraphics[width=\textwidth]{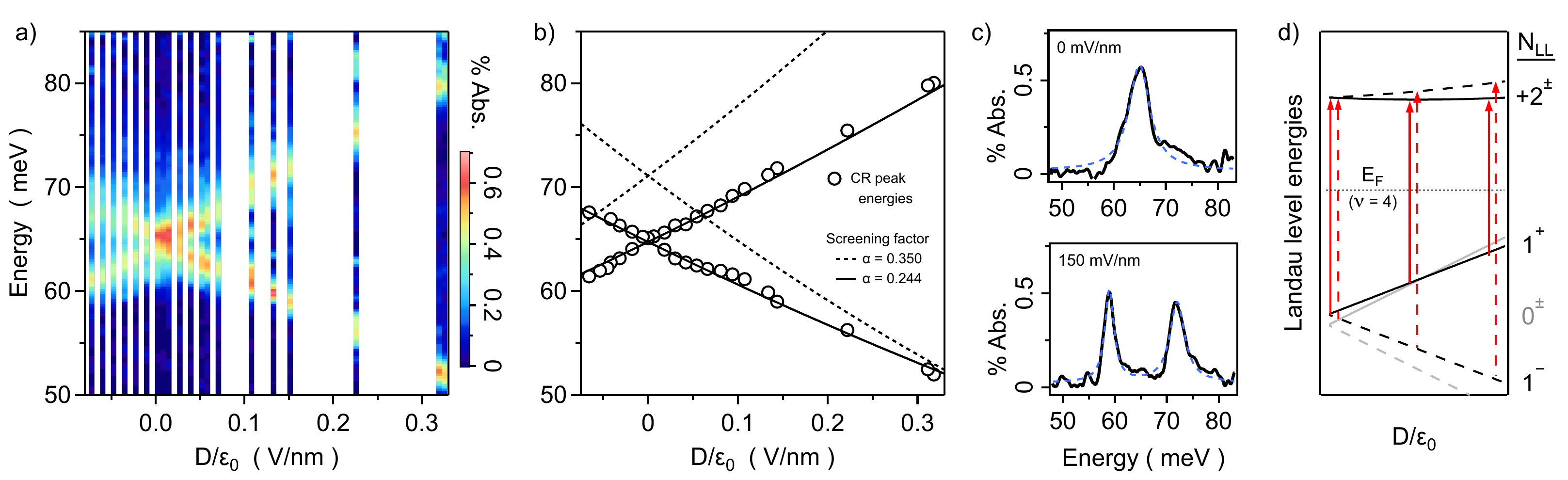}
\caption{(a) Color map of resonances at $\nu = 4$ vs displacement field, $D$. (b) Peak energies from fits to data in (a)\ \cite{SchosslerBLGfuture} (open circles) along with theoretically predicted transition energies (black lines) shown for two values of the screening parameter, $\alpha$. (c) Example spectra (black) for $D=0$ and $150$ mV/nm, with Lorentzian fits in dashed blue. (d) Landau level energies as a function of $D$; red arrows show allowed transitions between $N_{LL}=1$ and $2$, where superscripts mark the $K^{\pm}$ valleys. \label{nu4}}
\end{figure*}

We first orient this study on infrared transitions at fixed $\nu=+4$ where interaction effects are expected to be minimal. Figure \ref{nu4}(a) shows a color map of spectra for the $N=+1 \to +2$ LL transition at displacement fields $D/\epsilon_0$ from $-80$ to $330$ mV/nm; due to the long averaging times required for good signal to noise, spectra were not collected for all $D$ values. The peak transition energies are plotted in Fig.\ \ref{nu4}(b), as determined from Lorentzian curve fits to the data---examples are shown in Fig.\ \ref{nu4}(c)---or the location of peak maxima when curve fitting is inconclusive. The individual resonances are quite narrow, with line widths of 1-3 meV for individual resonances. Figure \ref{nu4}(b) also includes the results of our calculations to be described below. 

With reference to Fig.\ \ref{nu4}(d), at $\nu = +4$ the Fermi level lies between the $N=0,1$ octet and $N = +2$, so valley-conserving cyclotron transitions $\ket{N,\lambda} = \ket{1,+} \rightarrow \ket{2,+}$ and $\ket{1,-} \rightarrow \ket{2,-}$ are allowed where $\lambda=\pm$ denotes the $K^{\lambda}$ valley. The applied displacement field breaks the valley degeneracy of the $N = 0,1$ LLs (and $|N|\geq2$ as well, to a much smaller degree \cite{Zhang2011}) and generates an interlayer potential difference, $U = \alpha e D d/\epsilon_0$, with $d=0.34$ \AA\ the distance between layers, and $\alpha$ a phenomenological screening parameter. At the same time, transitions into the filled $N = 1$ level are blocked. Overall, one expects a single valley-degenerate resonance at $D = 0$ that splits in two for $|D| > 0$, with the splitting a direct measure of the induced valley gap in the $N = 1$ level. The data in Fig.\ \ref{nu4}(a,b) show a nearly linear-in-$D$ gap opening, similar to the zero field case \cite{Min2007,Zhang2009,Mak2009}.

We wish to know what level of theoretical approximation is required to accurately predict the $D$-dependence of the allowed transition energies shown in Fig.\ \ref{nu4}(d). We find very good agreement with these data using the theoretical model described above. At $\nu=+4$ this model has only a single adjustable parameter, $\alpha$, which determines the dimensionless screening strength. Interaction anisotropy has little impact far from charge neutrality and are disregarded for now. We adopt literature values for the SWMc terms \cite{Zhang2011}, and $\alpha$ is uniquely determined by its dual role in the model of both reducing the strength of Coulomb interactions and converting the applied electrical potential into a $D$-field. Therefore, it controls both the transition energies and the rate at which these energies change with the applied $D$. The data in Fig.\ \ref{nu4} are quite well fit by a single value of $\alpha = 0.244$ over the whole range of $D$, apart from a slight oscillation in the lower branch. For comparison, we also show curves calculated with $\alpha = 0.350$, which gives both a stronger $D$-dependence and shifts the energies up across the whole range of $D$. Beyond the screening term, this good fit is only achieved by including both many-body corrections to the single particle Landau level energies, and the more realistic SWMc model for bilayer graphene \cite{SchosslerBLGfuture}.

\begin{figure*}[t]
\includegraphics[width=\textwidth]{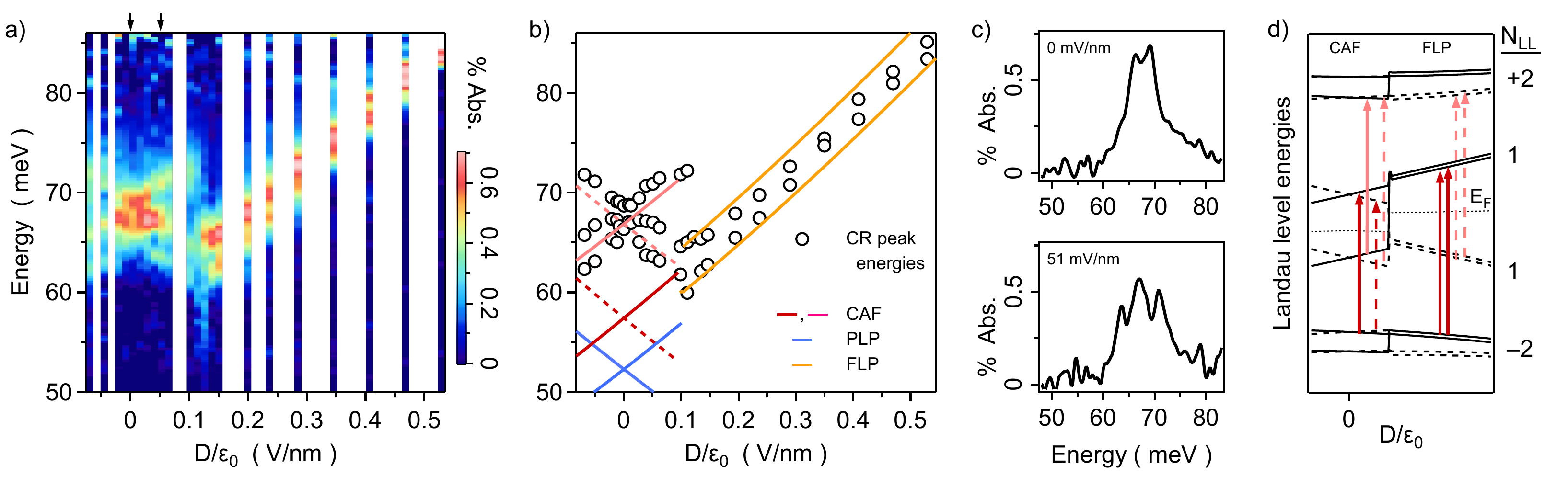}
\caption{(a) Color map of spectra at $\nu=0$ vs displacement field, $D$. (b) Resonance energies vs.\ $D$. Colorful lines are results of theoretical calculations for the canted antiferromagnet, partially layer polarized, and fully layer polarized ground states. (c) Example spectra showing the detailed line shapes for $D=0$ and 51 mV/nm, at the positions of black arrows in (a). (d) Calculated Landau level dispersion vs $D$, with dipole-allowed transitions $N=-2 \to+1~(+1\to+2)$ in red (pink). Transitions are shown  separately for both the CAFM and FLP regimes. The $N=0$ LLs are suppressed as they are not involved in these infrared transitions. \label{nu0} }
\end{figure*}

A color map of spectra measured at $\nu=0$ reveals further interesting and unexpected phenomena in Fig.\ \ref{nu0}. Line cuts like those shown in Fig.\ \ref{nu0}(c) reveal the fine structure due to multiple resonances, the energies of which are plotted in Fig.\ \ref{nu0}(b). Beginning precisely at $D=0$, two peaks of approximately equal weight can be resolved with a small splitting of 2.3 meV. These peaks increase in energy with differing slopes for increasing $D$, and a third lower-energy peak emerges above $D/\epsilon_0=26$ mV/nm. The weight of the middle peak is now roughly double the other two. With increasing $D$ these peaks disperse independently---the highest peak blue-shifts by a few meV while the lower two redshift with differing slopes---until, right at 120 mV/nm, all three peaks suddenly vanish to be replaced by a single broad peak with a long low-energy tail visible in Fig.\ \ref{nu0}(a). Immediately above this critical $D$ field, just two peaks reappear and follow a linear-in-$D$ dependence reminiscent of $\nu=+4$, except in lieu of a V-shaped splitting, both peaks increase in energy with a close approach or possible crossing near 350 mV/nm.  

As above, we seek to accurately calculate the energies of the allowed transitions shown in Fig.\ \ref{nu0}(d). Just like $\nu=+4$, at high $D$ the bilayer enters the fully layer polarized (FLP) regime where all charge is dragged to one side of the sample. The FLP prediction shown in Fig.\ \ref{nu0}(b) is a straightforward extension of the calculations above to the case of $\nu=0$, employing the same values of $\alpha$ and the SWMc parameters, and the results closely track the observed transitions. 

For weak displacement fields, however, there are four possible correlated ground state phases expected at $\nu=0$ in the model we employ, depending on the values of phenomenological short-ranged layer isospin anisotropy terms $u_z$ and $u_{\perp}$ \cite{Kharitonov2012}: a fully magnetized ferromagnet (FM); the FLP, and also a partially layer-polarized (PLP) state; and the canted antiferromagnet. Prior experimental work suggests the CAFM is the ground state for bilayer graphene \cite{Maher2013,Fu2021}. The FM is excluded since its quantized edge channels should lead to a finite spin Hall resistance rather than the gap seen in transport, and the FLP regime at high $D$ clearly does not extend below the sharp transition at $D/\epsilon_0=120$ mV/nm. Thus we calculate the energies of the allowed LL transitions $\ket{-2,\pm} \rightarrow \ket{1,\pm}$ and $\ket{1,\pm} \rightarrow \ket{2,\pm}$ for $D/\epsilon_0<120$ mV/nm, initiating the calculation by choosing values for the anisotropy terms which stabilize either the CAFM or PLP phases. The results are shown in Fig.\ \ref{nu0}(b).

Both the CAFM and PLP ground states lead to pairs of diverging linear-in-$D$ transitions arising as the valley symmetry is broken. Since the spin-degenerate PLP has only a single pair of transitions, vs two for the CAFM, we rule out the PLP phase because three transitions are plainly visible. The middle peak has roughly double the spectral weight of the others, implying it contains two nearly (perhaps accidentally) degenerate transitions for a total of four, consistent with the CAFM. 

Yet clearly there is only partial agreement between the data and the dispersions predicted for the CAFM mode. Most prominently, the CR peaks overlap only with the energy range of the upper set of CAFM transitions (pink in Fig.\ \ref{nu0}(b)), but not the lower (red). Also, the upper and lower resonances disperse with close to the expected slope vs $D$, but the middle peak shows only a weak and non-monotonic dependence on $D$; and the splitting at $\nu=0$ is unexpected.

\begin{figure*}[t]
\includegraphics[width=\textwidth]{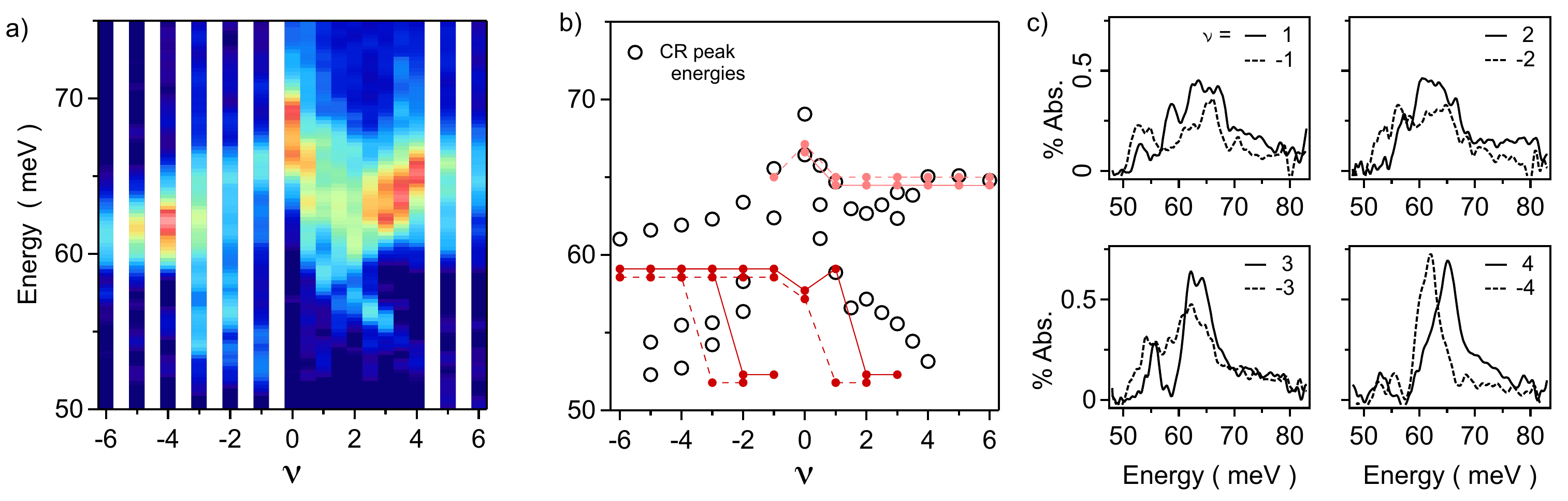}
\caption{(a) Color map of resonances observed at $D = 0$ vs filling factor, $\nu$. (b) Peak energies extracted from Lorentzian fits to the data. The red (pink) symbols are calculated transition energies for $N=-2\to+1 ~(+1\to+2)$, with solid (dashed) lines for the $K^+ (K^-)$ valleys. (c) Normalized spectra at some integer filling factors. \label{D0}} 
\end{figure*}

Finally, we study transitions at fixed $D = 0$ from $\nu = -6$ to +6 at integer fillings, and also at half-integers from $\nu = 0$ to $4$. A color map of the spectra is shown in Fig.\ \ref{D0}(a) with the peak energies plotted in Fig\ \ref{D0}(b). Multiple peaks disperse non-monotonically whose energies, spectral weights, and line widths have a moderate electron-hole asymmetry. The highest energies at $\nu=D=0$ are the same two peaks in the upper panel of Fig.\ \ref{nu0}(c). If these CR energies are converted to effective masses via $\omega_{c} = eB/m^{*}$, those in the upper branch ($>60$ meV) range from $m^*=0.022$ to $0.025 m_e$ ($m_e$ the free electron mass), and in the lower branch are $0.026-0.029m_e$. The latter approach the lightest masses extracted from compressibility or electronic transport data, while the electron-hole asymmetry of the upper branch is similar to prior trends at low density found in Shubnikov-de Haas oscillations \cite{Lee2014,Li2016}.

With the $B$ and $D$ fields held constant, the underlying single particle band structure is fixed so that the significant (${\sim}30\%$) variation in transition energies is entirely a consequence of electron interactions. We apply the same theoretical formalism as above, except the asymmetric terms are employed only for $|\nu| = 0, 1,$ and 2; at higher fillings their effect on the calculations becomes negligible. The outcome is quite good: gross features are readily captured including the overall highest energy at $\nu=0$, the electron-hole asymmetry, the presence of multiple branches, and the absolute range of energies. The agreement is not perfect, with details like the degree of asymmetry and the smooth decrease in transition energies with increasing $|\nu|$ not well reproduced.

In light of the good agreement of our calculations with the data for $\nu=4$, the conspicuous discrepancies for $\nu=0$ at low $D$, and across $D=0$, imply that despite the level of detail, the theoretical model fails to capture key phenomenology. 

For example, the unexpected splitting at $\nu=D=0$ in Fig.\ \ref{nu0}(c) may indicate coupling of the CR to a softened finite-$q$ magnetoroton mode \cite{perspectives,Zhao1995,Henriksen2006}. Fluctuations of the LL pseudospin ($N=0,1$) degree of freedom can result in nontrivial order including stripe-like states that break translational invariance and could drive the coupling \cite{YB2010, YB2012,Shizuya2020}; this ordering may be directly observable in scanned probe experiments. Separately, collective mode excitations (magnetoexcitons), not yet accounted for,  can induce CR shifts and splittings \cite{Kallin1984,iyengar_excitations_2007,Bychkov2008,Bisti2011,Shizuya2011,Toke2013a,Shizuya2020,Pack2020}.

The clear deviations in Fig.\ \ref{nu0} from the predictions for a straightforward CAFM to FLP phase transition may indicate the presence of interstitial phases between the CAFM at low $D$ and the FLP at high $D$ \cite{Lee2014,Hunt2017,Li2018}. In particular, the theoretical model we employ predicts the system should pass through the PLP phase intermediate between the CAFM and FLP \cite{Kharitonov2012}. Recent treatments also find several additional phases, particularly at high $B$ fields, with overlapping ground state energies at intermediate $D$ values; these include Kekule, broken U(1), and U(1)$\times$U(1) symmetries \cite{Knothe2016,Murthy2017,DeNova2017,Green2020}. 

If present, these phases will introduce first- and/or second-order phase transitions between the CAFM and FLP, broadening the range of $D$ over which the ultimate CAFM-to-FLP transition occurs. Since these potential phases are all insulators with a bulk gap and lack spin polarization, they allow all four non-degenerate transitions as the CAFM does. The overall lack of agreement with the CR peaks expected for the CAFM state calls for further theoretical work: we highlight a need to account for the impact of magnetoexcitons, explore the potential for LL pseudospin coherence, and predict signatures of potential alternative phases to the CAFM.

\bigskip

\begin{acknowledgments}
We thank K.\ Shizuya and J.\ Zhu for helpful correspondence. We acknowledge support for device fabrication and characterization from the Institute of Materials Science and Engineering at Washington University in St. Louis. JR acknowledges support from the Center for Quantum Leaps at Washington University in St.\ Louis. JB, YK, and EAH acknowledge support under NSF CAREER DMR-1945278. AS and MS acknowledge support under NSF Grant No.\ DMR-2029401.
\end{acknowledgments}

\bibliographystyle{apsrev4-1}

%\bibliography{bilayer.bib}

%

\end{document}